\documentclass[letterpaper, 10 pt, conference]{ieeeconf}  % Comment this line out if you need a4paper

\IEEEoverridecommandlockouts                              % This command is only needed if 
\usepackage{graphicx}
\usepackage{mathptmx} % assumes new font selection scheme installed
\usepackage{amsmath} % assumes amsmath package installed
\usepackage{pifont}
\usepackage{url}

\title{\LARGE \bf
Analyzing Gait Adaptation with Hemiplegia Simulation Suits and Digital Twins
}

\author{Jialin Chen$^{1}$, Jeremie Clos$^{2}$, Dominic Price$^{3}$ and Praminda Caleb-Solly$^{4}$% <-this % stops a space
\thanks{*This work was supported by the University of Nottingham, UK, and EPSRC EP/W000741/1}
\thanks{Author’s Accepted Manuscript. Released under the Creative Commons license: Attribution 4.0 International (CC BY 4.0) https://creativecommons.org/licenses/by/4.0/deed.enhttps://creativecommons.
org/licenses/by/4.0/}
\thanks{$^{1}$Jialin Chen is with School of Computer Science,
        University of Nottingham, UK
        {\tt\small psxjc16@nottingham.ac.uk}}%
\thanks{$^{2}$Jeremie Clos is with School of Computer Science, University of Nottingham, UK
        {\tt\small jeremie.clos@nottingham.ac.uk}}%
\thanks{$^{3}$Dominic Price is with School of Computer Science, University of Nottingham, UK
        {\tt\small dominic.price@nottingham.ac.uk}}%
\thanks{$^{4}$Praminda Caleb-Solly is with School of Computer Science, University of Nottingham, UK
        {\tt\small praminda.caleb-solly@nottingham.ac.uk}}%
}

\begin{document}

\maketitle
% \begin{center}
% \vspace{-2em}
% \footnotesize
% \textit{Author’s Accepted Manuscript. Released under the Creative Commons license: Attribution 4.0 International (CC BY 4.0) \\ \url{https://creativecommons.org/licenses/by/4.0/}}
% \vspace{1em}
% \end{center}

\thispagestyle{empty}
\pagestyle{empty}

%%%%%%%%%%%%%%%%%%%%%%%%%%%%%%%%%%%%%%%%%%%%%%%%%%%%%%%%%%%%%%%%%%%%
\begin{abstract}
To advance the development of assistive and rehabilitation robots, it is essential to conduct experiments early in the design cycle. However, testing early prototypes directly with users can pose safety risks. To address this, we explore the use of condition-specific simulation suits worn by healthy participants in controlled environments as a means to study gait changes associated with various impairments and support rapid prototyping.
This paper presents a study analyzing the impact of a hemiplegia simulation suit on gait. We collected biomechanical data using a Vicon motion capture system and Delsys Trigno EMG and IMU sensors under four walking conditions: with and without a rollator, and with and without the simulation suit. The gait data was integrated into a digital twin model, enabling machine learning analyses to detect the use of the simulation suit and rollator, identify turning behavior, and evaluate how the suit affects gait over time.
Our findings show that the simulation suit significantly alters movement and muscle activation patterns, prompting users to compensate with more abrupt motions. We also identify key features and sensor modalities that are most informative for accurately capturing gait dynamics and modeling human-rollator interaction within the digital twin framework.
\end{abstract}

\section{Introduction}
Maintaining mobility is crucial to ensuring independence and quality of life \cite{c21}, especially in post-stroke and individuals with frailty. Assistive aids such as rollators, accounted for over 67.2\% of the mobility assistance device market in 2021 \cite{c18}, highlighting their widespread use and importance. However, current assistive devices often lack automated adaptability, which limits their effectiveness in situations where patients' conditions change and there is a need for a system to adjust without expert intervention. Robotic technologies which can adapt to changing needs can play a significant role in rehabilitation \cite{c14} and supporting ambulation for activities of daily living \cite{c7}. Developing robotics devices which incorporate machine learning to enable adaptation require large amounts of representative data for training, particularly in the early stages of prototyping and testing. Digital twins (DTs) offer a promising solution to this challenge. A digital twin is a virtual replica of a physical system \cite{c2}, such as an assistive robot, that mirrors its real-world counterpart in real time. By capturing dynamic operational data through sensors, digital twins allow for continuous simulation, testing, and optimization of the system under various conditions. Unlike traditional simulations, digital twins are dynamic, with real-time feedback that enhances their accuracy and relevance in practical applications \cite{c17}. While digital twins have been widely adopted in manufacturing and healthcare \cite{c3}, their application in human-robot interactions, particularly in the design and testing of assistive robotics, remains under-explored.

This paper builds on the principles of digital twins to explore their ability to model and represent measures that relate to the interaction between a person and an assistive rollator. Specifically, it focuses on modelling and analyzing movement data collected from individuals walking with and without a rollator, and with and without a hemiplegia simulation suit \cite{hemiparesissuit}, using the Vicon system \cite{ViconSystem} and Delsys Trigno \cite{Delsys} surface EMG and IMU sensors. The study has three main objectives: to evaluate how effectively machine learning models can classify different walking conditions; to assess the impact of the hemiplegia simulation suit and rollator on movement and gait patterns; and to develop an adaptive digital twin capable of predicting a user's intention to turn based on sensor data.

In addition to achieving these objectives, we identify key sensors and features that effectively capture gait differences relevant to human–rollator interaction. This work serves as a step toward incorporating digital twin frameworks into the early-stage design and evaluation of assistive devices. Ultimately, it aims to reduce development time and support the creation of more responsive, user-centered technologies.

\section{Related Work}
\subsection{Digital Twins in Healthcare}
In the context of healthcare, DTs are increasingly used to model assistive technologies to enable experimentation with how best to personalize such systems to the needs of the user, allowing developers to test and refine designs without the need for expensive physical prototypes. For instance, Cascone et al. \cite{c4} developed a digital twin of a humanoid robot called Pepper and its surroundings, the simulated environment built was used to train the machine learning algorithms. Using a virtual environment helped to explore a range of human-robot interaction scenarios, gathering valuable feedback on different interaction design options. Similarly, the study by Jimene \cite{c12} et al. analyzed the impact of cyber-physical systems and digital twins on healthcare and showed how technology could lead to better patient care. Studies have demonstrated the potential of DTs to improve the safety and functionality of assistive devices like rollators \cite{DTsAssistive}, ensuring they meet the unique requirements of individuals with mobility issues.

\subsection{Smart Walkers and Sensor Configuration}
Smart walkers, equipped with sensors, motors, and actuators, aim to improve mobility by offering features like obstacle detection and assistive navigation \cite{c16}. The LEA robotic walker \cite{c9}, for instance, was a smart rollator equipped with numerous sensors designed to assist mobility, particularly for patients with Parkinson's, however failed to gain market traction, with an average price of £8.5k when it was available for sale. In order to ensure wide-scale adoption within resource constrained health and social care settings, intelligent robotic interventions need to be easily accessible and affordable to many who could benefit from their advanced features, but making the right decisions regarding the optimal minimal-viable system to maintain economic sustainability is a challenge. There is an opportunity for conducting more testing in simulation to determine the best minimal configuration of sensors that responds to changing user needs, and hence ensure economic viability.

\subsection{Biosignals and Motion Data for Gait Analysis}
Reliable data from gait monitoring and tracking is crucial for designing adaptive walkers. Inertial Measurement Units (IMUs) and Electromyogram (EMG) sensors, often used alongside motion capture systems like Vicon, provide valuable data for gait analysis. Mao et al. \cite{c15} demonstrated the effectiveness of using surface electromyography(sEMG) and acceleration (ACC) signals for predicting grip force and wrist angles using support vector machines (SVM) \cite{SVM}. This example highlights the capability of EMG-based analysis to potentially capture the force being applied by a person on the handlebars as they use a rollator.
\subsection{Machine Learning and Data Handling}
Using machine learning within biomechanical analysis can simplify the complex interpretation of gait parameters. Supervised Machine Learning (ML) techniques, particularly Support Vector Machines (SVMs), have achieved over 90\% accuracy in gait analysis \cite{c13}. Deep learning approaches, such as Convolutional Neural Networks (CNNs) \cite{c6} and long short-term memory networks (LSTMs) \cite{c5}, have shown promise in human activity recognition and gait stability prediction. Data handling methods, including band-pass filtering for EMG \cite{filter} and low-pass filtering for IMU data, are commonly used to reduce noise. Time-based segmentation helps structure the data, while handcrafted features, such as RMS, zero-crossings, and signal magnitude area, are frequently extracted to capture key signal characteristics and improve model performance \cite{feature}.

\section{Proposed Digital Twin system}

\subsection{Physical Layer}
\begin{figure}[htbp]
    \centering
    \includegraphics[width=.8\linewidth]{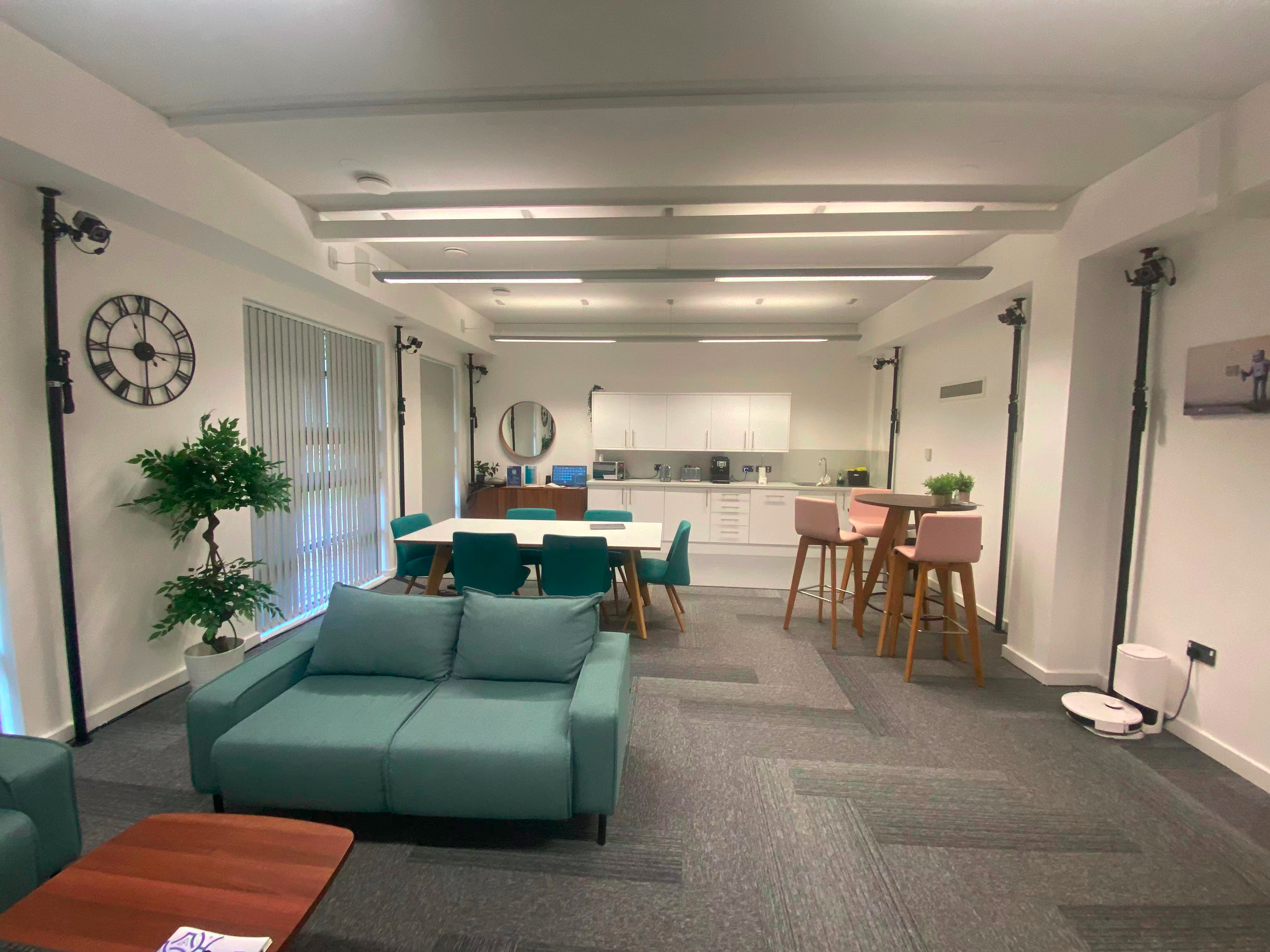}
    \caption{Vicon Vero cameras and the living lab environment}
    \label{fig:livingroom}
\end{figure}

The physical layer consists of components in the real-world environments, assistive devices (e.g., a rollator) and sensors integrated into the device, as well as on the user and their surroundings. This layer represents the direct interactions of the human users, capturing their movements through physiological and motion capture sensors. In the first phase of our research we conducted data collection with healthy participants in a controlled environment (Fig. \ref{fig:livingroom}), using a Vicon Nexus motion capture system with eight Vicon Vero cameras, two RGB cameras, and Delsys Trigno sensors. 
   
\begin{figure}[htbp]
    \centering
    \includegraphics[width=.95\linewidth]{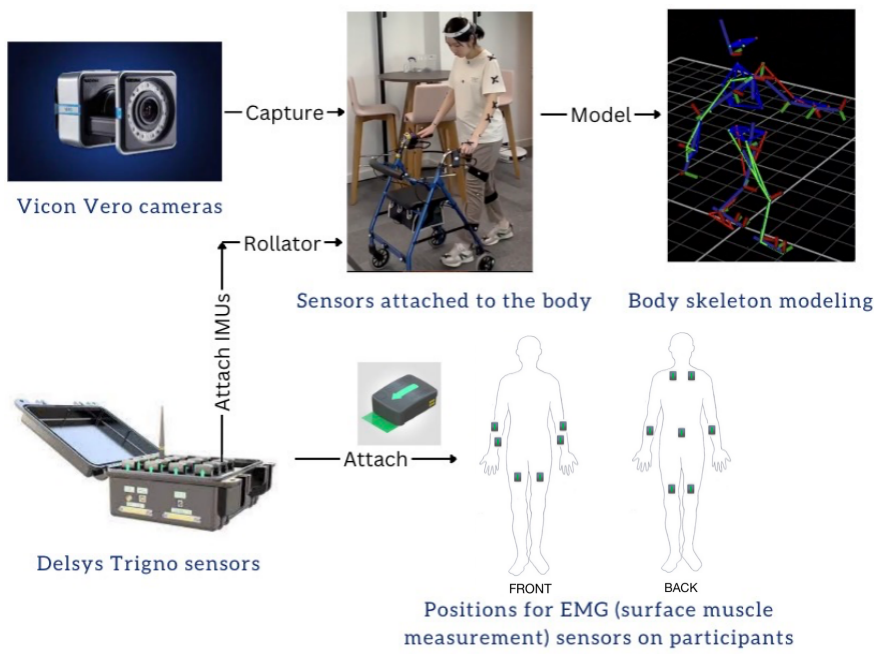}
    \caption{Integrated Vicon Nexus system and Delsys Trigno system for human biomechanical modeling by attaching sensors on the body}
    \label{fig:sensor_flowchart}
\end{figure}

Vicon and Trigno sensors (Fig. \ref{fig:sensor_flowchart}) were placed on the participants' to monitor gait, posture, and other biomechanical parameters, in an environment that replicated a home studio apartment.

Data was collected in four different scenarios, which involved participants walking with and without a rollator, as well as with and without a hemiplegia simulation suit. This suit \cite{hemiparesissuit} included a knee splint, a weight cuff, an eye patch, an elbow wrap, ear defenders, and ear plugs, simulating some of the physical limitations resulting from hemiplegia. The data collected during these tasks provided critical insights into how physical impairments affect gait and movement patterns. This data was essential for training our digital twin models to recognize and predict user intentions in the virtual layer.
% \begin{figure}[htbp]
%     \centering
%     \includegraphics[width=.7\linewidth]{Figures/suit1.png}
%     \caption{Simulation suit with an elbow wrap, an ear defender, an eye patch, a pair of earplugs, a knee splint, and a weight cuff used in the study}
%     \label{fig:suit}
% \end{figure}

\subsection{Virtual Layer}
\begin{figure}[htbp]
    \centering
    \includegraphics[width=.95\linewidth]{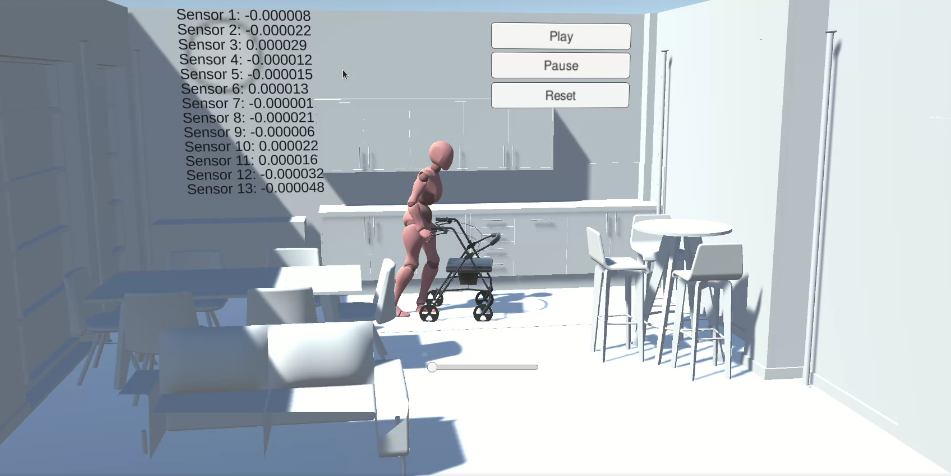}
    \caption{Virtual space with entity models and sensor data stream}
    \label{fig:virtualworld}
\end{figure}

The virtual layer represents the digital counterpart of the physical system, integrating detailed simulations of the user's biomechanics, the assistive device, and the environment (Fig. \ref{fig:virtualworld}). 

This layer is intended to serve as a platform for real-time analysis and visualization of human-rollator interactions. The prediction of human actions based on collected data can also be presented in this layer. Furthermore, human and rollator characteristics can be explored by updating the variables in the system. Virtual entity modeling was explored in MuJoCo \cite{mujoco} and Unity \cite{unity}, each entity model with physical properties was integrated into a virtual environment and led to a digital twin system with different modes. One is the recorded mode, using the collected data to map the positions and interactions between real human-rollator movements, and the other is the virtual one, with playback functions to update variables (such as flooring, uneven surfaces with trip hazards, height of rollator, balance of humans, etc.) in the system to figure out how changing parameters affect human and rollator characteristics. Another mode is the real-time one, with real-time sensor data flowing into the trained machine learning models to provide feedback from the rollator to users. 

\subsection{Data Layer}
The data layer is the core driver of the DT system, connecting the physical and virtual layers and managing the flow of data between them. This layer is responsible for storing, processing, and managing the vast amounts of data collected from human participants. 

The data layer supports both offline and real-time operations:

Offline Data Learning: in this mode, data collected from participants, such as EMG signals and dynamic movement, through acceleration and gyroscope signals, were used to train machine learning models. These models detect patterns in gait and body movements, facilitating the prediction of user intention, as they prepare to turn. In this paper we present results of how the system is used to assess the impact of wearing the hemiplegia simulation suit on gait characteristics, helping to refine the models for real-world application. Once trained, these models can be tested within the virtual layer, and their feedback can be later optimised for deployment in real-world scenarios.

Real-Time Data Processing: in the real-time mode, sensor data is continuously streamed from the physical layer to the data layer. The models trained from the recorded mode (off-line data learning) will process and analyze the data to detect anomalies or patterns that may indicate a need for intervention. For instance, if a user did not show signs of turning away while walking straight towards an obstacle, the system could help avoid a potential collision by sending feedback to the user via the physical layer 
as an alert to avoid a potential collision or fall, with the assistive device adapting its behavior dynamically based on the user's real-time status.

\begin{figure}[htbp]
    \centering
    \includegraphics[width=\linewidth]{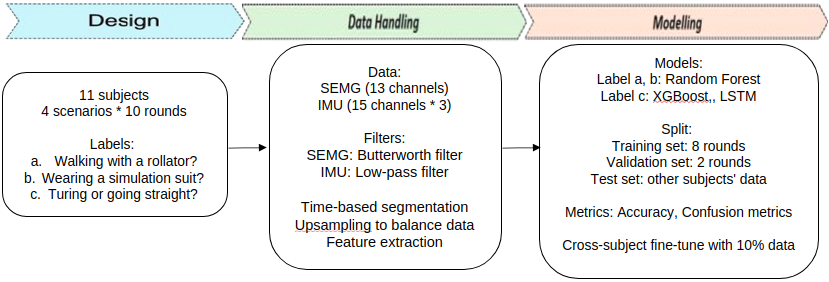}
    \caption{Machine learning based pipeline for human-rollator interaction scenario}
    \label{fig:mlpipeline}
\end{figure}

In both of the data processing modes, machine learning approaches will be applied to make classifications and predictions. A machine-learning based pipeline in our research is shown in Fig. \ref{fig:mlpipeline}, from data collection, processing, to modeling, as explained in the following sections.

\section{Experimental Setup}
\subsection{Participants}
The first phase of data collection involved 11 healthy participants [mean age: 27; gender: 5 female, 6 male]. Ethical approval for the study was obtained from the Computer Science Ethics Committee (Ref no. CS-2022-R50). To ensure that the study design met the needs and concerns of end-users, a consultation with members of the National Rehabilitation Centre Patient and Public Involvement and Engagement (NRC PPI) group was held.

\subsection{Procedure}
Participants were screened using an online questionnaire (https://forms.office.com/e/k5XGRRZtEd) to ensure their suitability for the study. Those who passed the screening were asked to complete an online demographic questionnaire and were then invited to the Cobot Maker Space's living lab at the University of Nottingham.

Upon arrival, participants were re-screened and gave their written consent before the data collection began. Each session lasted approximately 1.5 hours. The rollator height was adjusted for each participant based on their wrist level while standing, as recommended by the literature \cite{c19}. Next, 39 reflective markers were attached to the participant's body according to the standard full-body Plug-in Gait model \cite{c22}, and anthropometric measurements such as participant's height, weight, knee width, joint length, etc. were recorded for use in Vicon Nexus body modeling.

\begin{figure}[htbp]
    \centering
    \includegraphics[width=.8\linewidth]{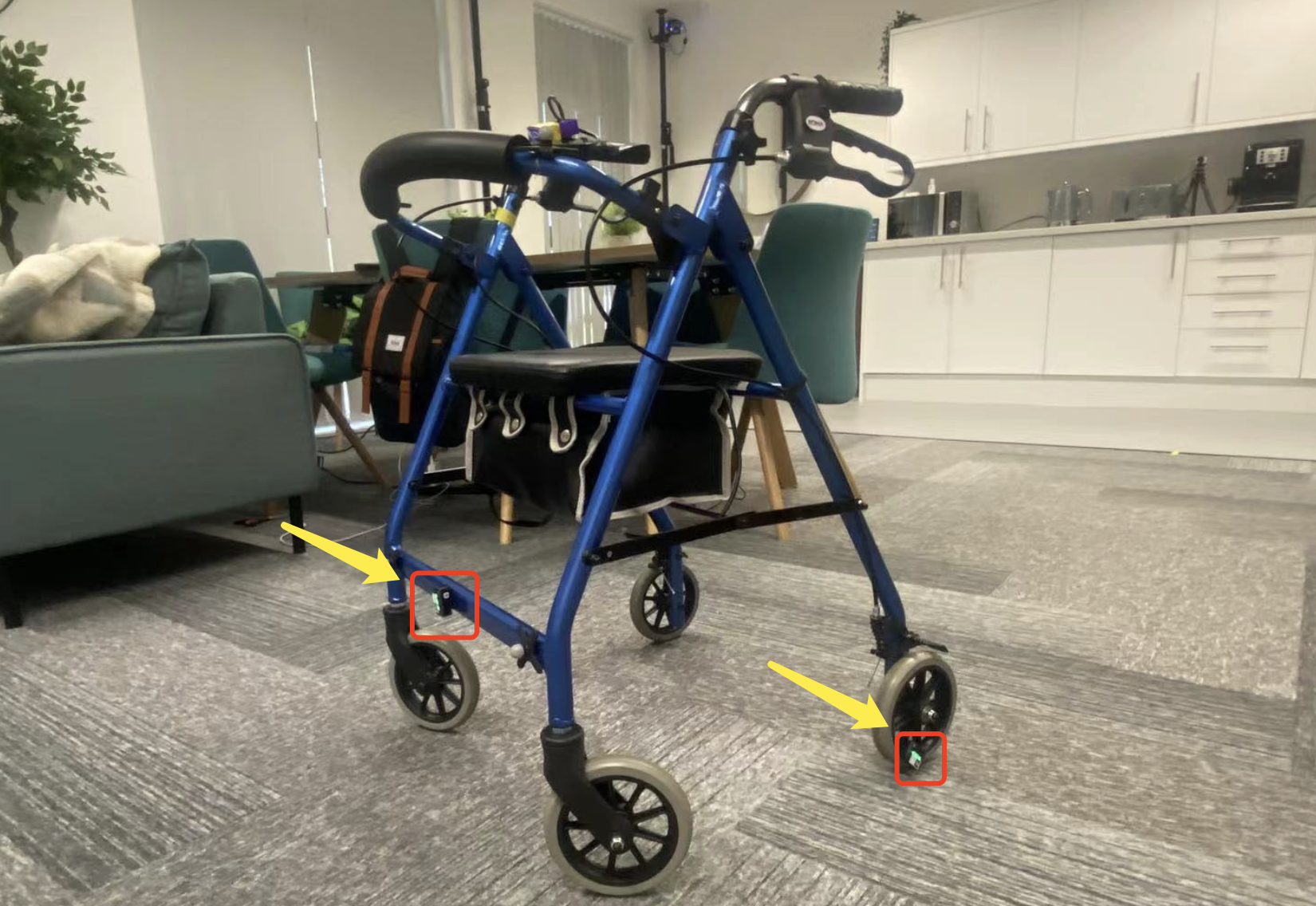}
    \caption{Positions for the Delsys sensors on the rollator}
    \label{fig:rollator}
\end{figure}

Additionally, 13 Delsys Trigno sensors were placed on key muscle groups to collect sEMG data, with sensors attached over the rectus femoris, biceps femoris, brachioradialis, flexor digitorum superficialis, extensor carpi ulnaris, erector spinae, and trapezius muscles \cite{c1, c8, c11}. Two additional sensors were attached to the rollator (Fig. \ref{fig:rollator}) to capture acceleration and gyroscope data.

Participants performed a static "motorbike" pose for 2-3 seconds to calibrate the Vicon system, ensuring accurate tracking of the reflective markers and the generation of a labeled skeletal model.
% \begin{figure}[htbp]
%     \centering
%     \includegraphics[width=.8\linewidth]{Figures/motorbike.png}
%     \caption{Human subject static calibration in Vicon system, from reflective markers to human skeleton setup}
%     \label{fig:motorbike}
% \end{figure}

\begin{figure}[htbp]
    \centering
    \includegraphics[width=.8\linewidth]{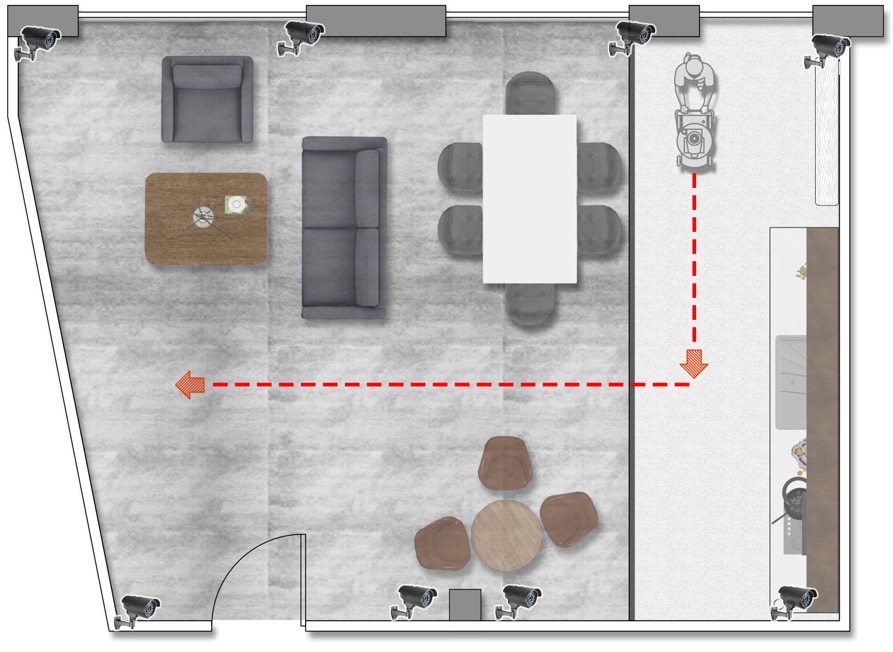}
    \caption{Layout of the living lab with experimental route}
    \label{fig:route}
\end{figure}

The participants then completed 10 rounds of walking along an "L" shaped path and back, for each scenario in the lab (Fig. \ref{fig:route}), under four different conditions: with and without the rollator, with and without the simulation suit on the left or right side of their body (simulating hemiplegia). 

\subsection{Data Collection}
The initial phase focused on collecting baseline data from healthy participants, in four scenarios (Table \ref{tab:scenario}) to model human movement and assess mobility. The simulation suit was used to simulate hemiplegia, a common post-stroke condition, enabling comparison between healthy participants and future data collection from actual patients with mobility impairments.
\begin{table}[htbp]
  \caption{Four scenario descriptions for the data collection from healthy participants}
  \centering
  \label{tab:scenario}
  \begin{tabular}{ccl}
    \hline
    Scenario & Walking with a rollator? & Wearing the simulation suit?\\
    \hline
    1 & No & No\\
    2 & No & Yes\\
    3 & Yes & No\\
    4 & Yes & Yes\\
    \hline
\end{tabular}
\end{table}

% \begin{figure}[htbp]
%     \centering
%     \includegraphics[width=.8\linewidth]{Figures/sequence.png}
%     \caption{Temporal layout of the designed experiment}
%     \label{fig:sequence}
% \end{figure}

Participants walked along a predefined path while wearing the simulation suit on either the left or right side, reflecting the nearly equal prevalence of left- and right-hemispheric strokes \cite{c10}. This design allows for a direct comparison between simulated and real-world impairments in future phases. The collected data included 1-channel sEMG data at 2 kHz from each Trigno sensor and 6-channel acceleration and gyroscope data at 200 Hz from the rollator sensors.
% An example of full-body modeling and real-time sEMG data from a participant wearing the simulation suit and using the rollator is shown in Fig. \ref{fig:fullbody}.
% \begin{figure}[htbp]
%     \centering
%     \includegraphics[width=.8\linewidth]{Figures/skeletonwithEMG.png}
%     \caption{Full body modeling using Plug-in Gait template from a subject wearing the simulation suit and walking with the rollator, with sEMG data showing on the right from the Trigno sensors}
%     \label{fig:fullbody}
% \end{figure}

\section{Data Processing and Modeling}
\subsection{Data Exploration and Visualization}
\begin{figure}[htbp]
    \centering
    \includegraphics[width=1\linewidth]{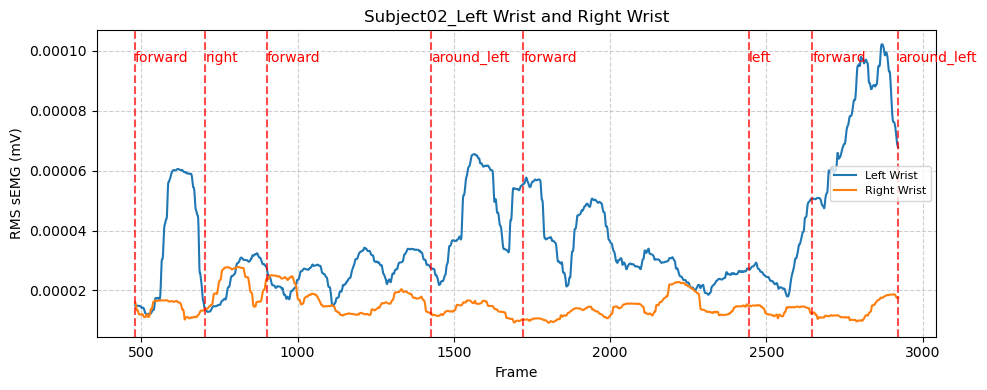}
    \caption{Extracted RMS feature from left wrist and right wrist muscle intensity over the first trial in subject 02 walking with a rollator}
    \label{fig:wrist}
\end{figure}

\begin{figure}[htbp]
    \centering
    \includegraphics[width=1\linewidth]{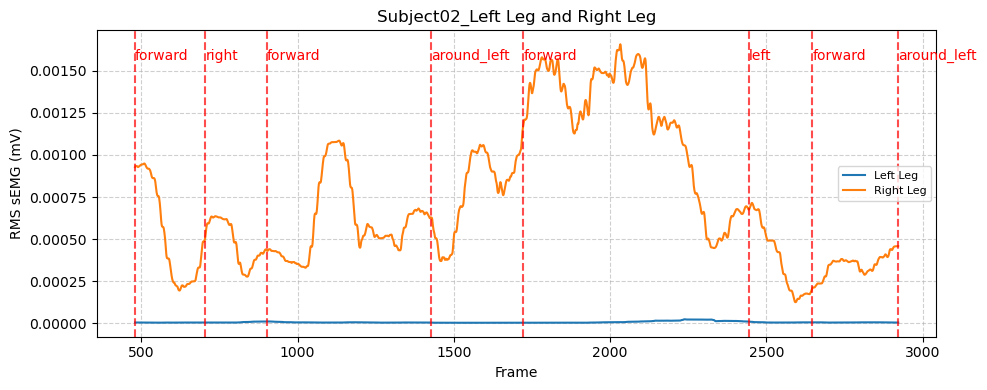}
    \caption{Extracted RMS feature from left leg and right leg muscle intensity over the first trial in subject 02 walking with a rollator}
    \label{fig:leg}
\end{figure}

Missing motion capture data were interpolated using a polynomial method to fill gaps caused by sensor occlusion or error. A bandpass filter was applied to each sEMG channel using a Butterworth filter with a low cut-off of 20 Hz and a high cut-off of 450 Hz. An example of the muscle force intensity over time for a walking trial of subject 02 with a left hemiplegia suit in the scenario of walking with a rollator is shown in Fig. \ref{fig:wrist}. The Root Mean Square (RMS) value was calculated and extracted for each channel to show muscle activation intensity during movements. Sensors were grouped by five body segments according to the attached positions: back, left wrist, right wrist, left leg, and right leg. Fig. \ref{fig:wrist} plots the changes in the force of the left wrist and right wrist over time, and the force of the left wrist muscle is higher than that of the right wrist muscle intensity. On the contrary, Fig. \ref{fig:leg} shows right leg muscle intensity is higher than left leg muscle intensity. When comparing subjects with opposite hemiplegia side, it was found that the non-restricted leg muscle generates more force when the opposite wrist is immobilized to compensate and help with balance.

\subsection{Classification Tasks Using the Collected Data}
To further analyse the effect of how the hemiplegia suit and rollator affect muscle intensity and movements, two classification tasks were defined as noted in the following sections.
\subsubsection{Detection of Simulated Hemiplegia and Rollator Usage} 
this task assessed the presence or absence of external suit constraints and rollator affecting movement. Traditional classifiers were employed. Between subjects and within subject analysis were performed. For inter-subject analysis, cross-validation followed a leave-one-subject-out strategy. For intra-subject analysis, cross-validation followed a leave-two-rounds-out strategy, where rounds were defined by experimental sessions. This ensured models generalized across different test sets and conditions. Accuracy was calculated to show model performance.

\subsubsection{Detection of Movements} 
\begin{figure}[htbp]
    \centering
    \includegraphics[width=.8\linewidth]{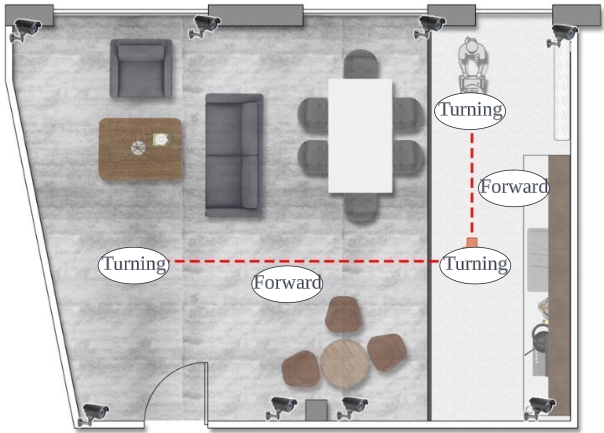}
    \caption{Labels showing the turning and forward classes}
    \label{fig:turnforward}
\end{figure}

in this classification task, five primary movements (forward, turn left, turn right, turn around from left, and turn around from right) performed by participants in the lab were extracted. In this task, all turns (90° turns and 180° turns) are included in the ``turning" class shown in Fig. \ref{fig:turnforward}, and others are ``forward" class. Cross-validation and performance metrics follow the same principle as for simulated hemiplegia and rollator detection. Our hypothesis is that if we are able to accurately classify the different movements then we could use this information to automatically adjust the rollator, as well as provide pro-active feedback to support users better in the future.

\subsection{Data Processing and Feature Extraction}
Data Preprocessing: missing values were handled through imputation, and outliers were detected and removed based on predefined thresholds. EMG signals were bandpass-filtered (20–450 Hz) to remove noise and retain meaningful muscle activation frequencies. Segmentation was performed using a time-based dynamic sliding window, with a window size of 400 samples (200 ms at 2000 Hz) and a step size of 200 samples (100 ms), ensuring 50\% overlap to effectively capture temporal transitions.

Feature Extraction: key features shown in Table \ref{tab:feature} were extracted from sEMG values \cite{emgfeature} and IMU signals \cite{imufeature}. 

\begin{table}[htbp]
  \caption{Extracted features from Trigno sensors}
  \centering
  \label{tab:feature}
  \begin{tabular}{cccl}
    \hline
    Feature & sEMG & Acc & Gyro\\
    \hline
    Root Mean Square & \ding{51}     & \ding{51}    & \ding{51}\\
    Variance & \ding{51}     & \ding{55}    & \ding{55}\\
    Mean & \ding{55}     & \ding{51}    & \ding{51}\\
    Standard Deviation & \ding{55}     & \ding{51}    & \ding{51}\\
    Jerk & \ding{55}     & \ding{51}    & \ding{51}\\
    Simple Moving Average & \ding{55}     & \ding{51}    & \ding{51}\\
    Mean Absolute Value & \ding{51}     & \ding{55}    & \ding{55}\\
    Slope Sign Changes & \ding{51}     & \ding{55}    & \ding{55}\\
    \hline
\end{tabular}
\end{table}

\section{Results}
\subsection{Detection of Simulated Hemiplegia and Rollator Usage}
Data collection was divided into 4 scenarios of wearing, without wearing a hemiplegia simulation suit, walking with and without a rollator from 11 subjects. A generalized model was trained using 10 subjects' EMG data, 1 subject's data was extracted as test set. While each person's muscle activation and movement patterns is unique, the generalized model accuracy had a poor performance. So personalized models were trained instead of the generalized one. By splitting 2 rounds as test set and 8 rounds as training set for each scenario in every subject, 2 classifiers were trained separately for suit detection and rollator detection, both models captured individual movement patterns very well (99\% average accuracy on the test set) for every subject. The high accuracy shows impaired gait from healthy participants wearing a hemiplegia simulation suit and normal gait, walking with and without a rollator can be reliably distinguished when doing intra-person modeling.

\subsection{Adaptation Analysis}
\subsubsection{How Does the Suit Affect Movement}
\begin{figure}[htbp]
    \centering
    \includegraphics[width=.8\linewidth]{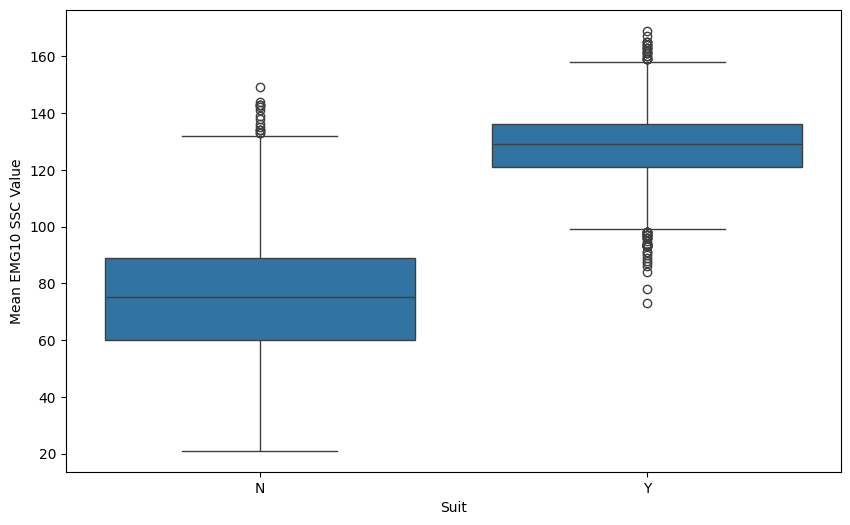}
    \caption{Mean values of feature SSC from sensor 10 (positioned at left leg) when not wearing the suit and wearing the suit}
    \label{fig:featureMean}
\end{figure}

after the data processing and feature extraction stages, the mean value was calculated for each feature. We found that the Slope Sign Changes (SSC) feature is the most sensitive to whether the subject was wearing the suit or not. Fig. \ref{fig:featureMean} is an example to show how the suit affects movement by comparing the mean values of the different features. The example in the figure shows the difference between when a subject is wearing the suit and their left leg is restricted, and the values of the feature when they are not wearing the suit. It is noted that the suit significantly alters muscle activation patterns, especially for feature SSC, which increases abrupt muscle activation changes after wearing the suit that restricted movement. Less affected Root Mean Square features indicates that the suit forces the user to compensate with more abrupt movements, rather than smooth ones. Similar trends were found in other subjects.

% The differences between left and right sides for the same subject was plotted in Fig. \ref{fig:dif}, sensor 11 was on left leg and sensor 13 was on right leg, legs were the most affected parts by the suit. From the figure, the suit reduces left-right muscle activation difference in the legs but introduces high variability (outliers), it was concluded that while the suit restricts movement, it might be affecting control inconsistently across time.
% \begin{figure}[htbp]
%     \centering
%     \includegraphics[width=.7\linewidth]{Figures/dif.png}
%     \caption{Differences from left side and right side, sensor 4, 5, 6 were on left wrist, sensor 7, 8, 9 were on right wrist, sensor 10, 11 were on left leg, ensor 12, 13 were on right leg}
%     \label{fig:dif}
% \end{figure}

\subsubsection{Trends Over Rounds}
\begin{figure}[htbp]
    \centering
    \includegraphics[width=.8\linewidth]{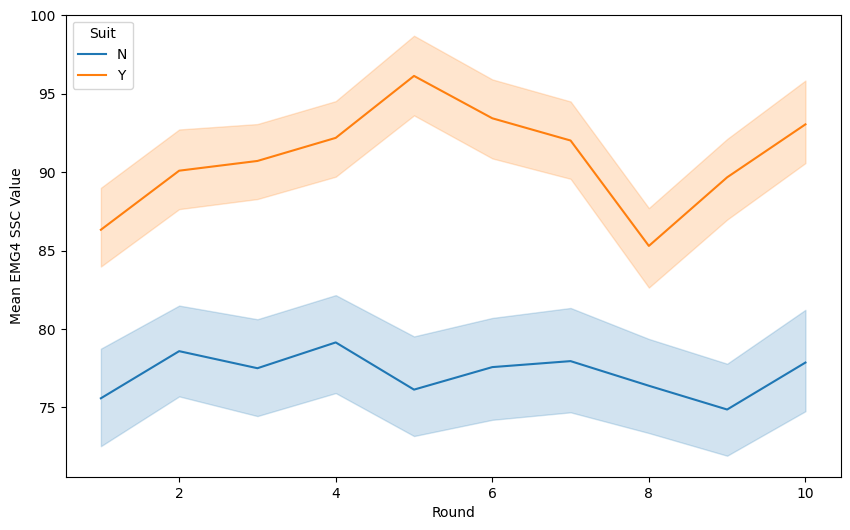}
    \caption{Trends over rounds for one subject as an example}
    \label{fig:trend}
\end{figure}

each subject walked 10 rounds in every scenario, line plots were drawn to find whether there was a correlation with adaptation and fatigue over time. One example of a subject trend was in Fig. \ref{fig:trend}, similarily for other subjects. From visualization, there is no clear trend in adaptation over rounds using sEMG data, it seems the suit imposes a continuous restriction. 

\subsection{Movement Classification}
\begin{figure}[htbp]
    \centering
    \includegraphics[width=.95\linewidth]{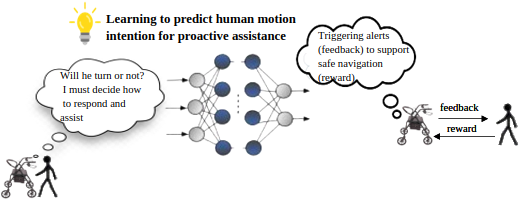}
    \caption{Human motion intention to assist walking}
    \label{fig:int}
\end{figure}

Turning intention detection (Fig. \ref{fig:int}) predicts whether the user intends to turn or not, especially in front of an obstacle, enabling the rollator to proactively trigger vibration or voice alerts to avoid falls and enhance safe navigation.

\begin{figure}[htbp]
    \centering
    \includegraphics[width=.8\linewidth]{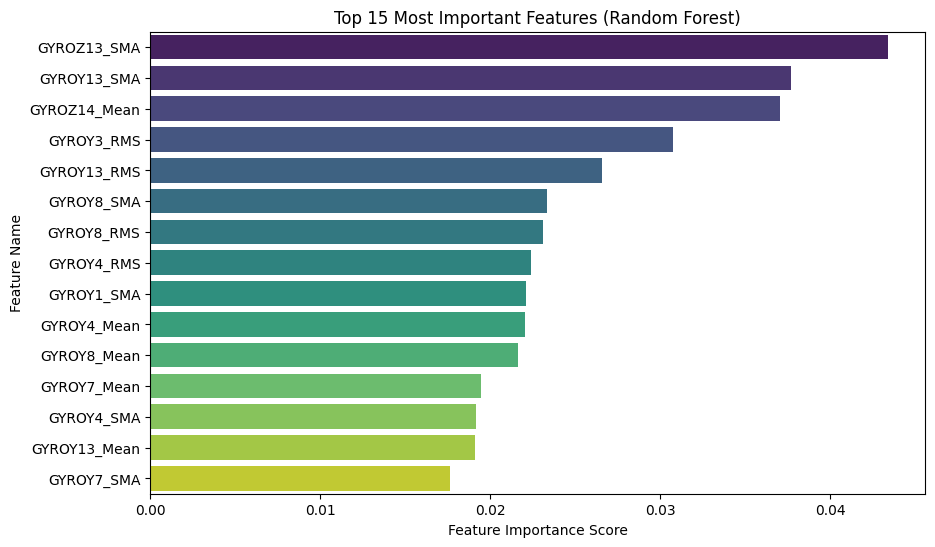}
    \caption{Most important features extracted using RandomForest model}
    \label{fig:imp}
\end{figure}

In our datasets, movements were grouped as ``walking straight" and ``turning". After processing IMU signals described in the last section, feature importance was shown in Fig. \ref{fig:imp} from Random Forest model. Most important features are from Gyro signals, SMA and RMS are key features to turning.

\begin{figure}[htbp]
    \centering
    \includegraphics[width=.8\linewidth]{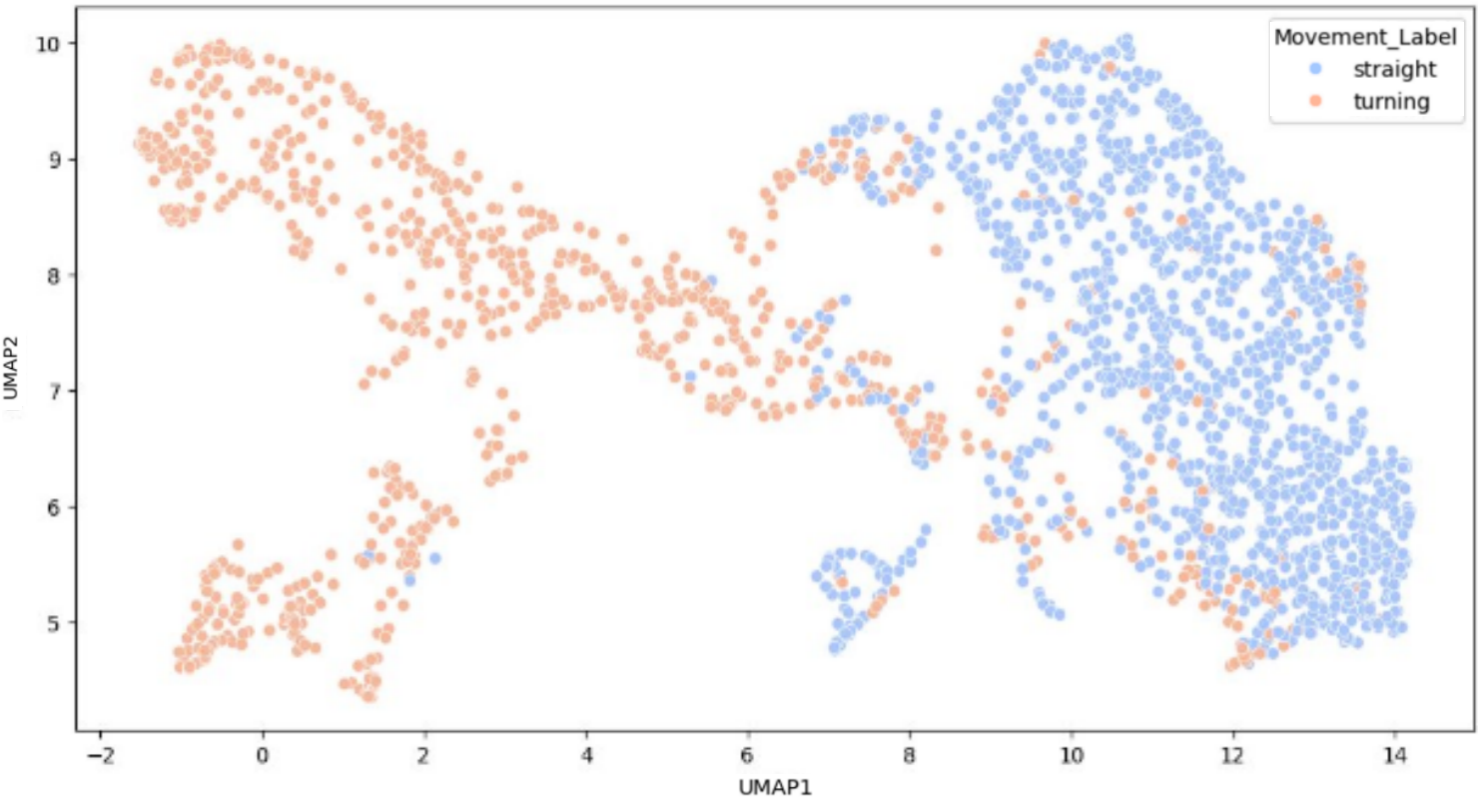}
    \caption{UMAP visualization of feature data}
    \label{fig:um}
\end{figure}

 UMAP \cite{UMAP} was used to reduce dimensionality and explore relevance, from Fig. \ref{fig:um}, a distinction can be found roughly for the movement labels. 

After data handling, XGBoost\cite{XGBoost} was trained to classify the movements for one subject by leave-two-rounds-out cross validation strategy achieving an average accuracy of 93.5\%. Using the trained XGBoost model and applying it to other subjects, the average accuracy dropped to 75\%. By using 10\% of a new subject's data to retrain the model, and testing on the remaining 90\%, average accuracy increased to 86\%. 
% Fig. \ref{fig:cm_sgb} shows a confusion matrix of the XGBoost model re-trained with 10\% data from a new subject and tested with the remaining 90\%, 0 is forward class, 1 is turning class.
% \begin{figure}[htbp]
%     \centering
%     \includegraphics[width=.56\linewidth]{Figures/cm_xgb.png}
%     \caption{Confusion matrix for turning and forward classes }
%     \label{fig:cm_sgb}
% \end{figure}

LSTM was also trained with the training and test partition of the data and compared with XGBoost, and reached 89\% accuracy for the test set. It was suggested that the trained model can be adapted to a new subject by incorporating a small portion of the new subject’s data for fine-tuning or retraining, which significantly improves generalization performance across subjects.

\section{Discussion}
This study considered the impact of a hemiplegia simulation suit on movement patterns using Vicon motion system, Delsys Trigno EMG and IMU data to capture the dynamics of the gait. The increased values of Slope Sign Changes indicated abrupt muscle activation when wearing the suit, suggesting a compensatory mechanism to help maintain balance. These findings validate the efficacy of the simulation suit in replicating hemiplegia gait restriction characteristics, making it a valuable tool for prototyping and testing assistive devices in controlled environments.

The combination of EMG and IMU data provided a robust feature set for walking conditions detection and classification demonstrating the value of multi-sensor fusion in mobility analysis. Additionally, turning movements were successfully classified with 89\% accuracy, which shows potential for real-time prediction of turning movement intentions. In future work we hope to use trained machine learning models to provide actionable reminders (such as physical vibration alarms, voice alarms etc.), by integrating the collected data and models to the digital twin system to test model functions and rollator functions. 

Our results contribute to the ongoing development of digital twin frameworks for assistive technologies, enabling more agile and personalised modelling of human-device interactions, and in turn inform the design and early testing of adaptive assistive devices, such as intelligent rollators. By integrating real-time biomechanical data with predictive, assistive technologies could dynamically adjust support parameters, responding to changing gait and user needs.

\end{document}